**To cite this article:** P. Kusawat and N. Rompho, "Impact of Investing Characteristics on Financial Performance of Individual Investors: An Exploratory Study," *2019 IEEE International Conference on Industrial Engineering and Engineering Management (IEEM)*, Macao, China, 2019, pp. 654-658, doi: 10.1109/IEEM44572.2019.8978725.

**To link to this article:** https://doi.org/10.1109/IEEM44572.2019.8978725

P. Kusawat[1], N. Rompho[2]
[1]Ph.D. Student, Department of Marketing, Thammasat University, Bangkok, Thailand
[2] Professor, Department of Operations Management, Thammasat University, Bangkok, Thailand
[1]poompak-kus61@tbs.tu.ac.th, [2]nopadol@tbs.tu.ac.th

*Abstract* - **This exploratory study examines which investing characteristics determine success in an equity market. Based on data from 403 respondents, exploratory factor analysis results in 13 factors: middle/long time horizon, qualitative analyst, open-minded/disciplined, organized, emotional stability, naïve, growth stock, concentrated portfolio, contrarian, value stock, globalized, intrinsic value, and price-independent. Multiple linear regression of individual investors' excess return on these factors show statistically significant relationship. These results deepen our knowledge on what sort of investing characteristics are required to survive in equity markets.**

*Keywords* – **investing characteristics, individual investors, exploratory factors analysis**

## I. INTRODUCTION

This study examines which characteristics of an investor can determine success in an equity market. The term "investor" in this paper is used in a generic sense to refer to individuals who engage in trading stocks for their own accounts. This study does not explicitly distinguish between different types of investors.

The argument that investors' characteristics could possibly affect their performance is a challenge towards the proposition in traditional finance. Largely based on Efficient Market Theory [1, 2] and Modern Portfolio Theory [3], it is advocated that investors are rational and base their decisions on all relevant publicly available information. Since all these rational market agents make informed decisions in unbiased ways, all investors in the same market exposed to the same set of information would hold identical portfolio, in effect, prices would fully be reflecting all available information at all times. However, some research findings express doubt about the rationality of investor behaviors, empirical evidence in the behavioral finance literature shows that individuals do not behave rationally [4], supporting the view that markets are not efficient and investor biases affect security prices substantially. Prior research has also proven that investors tend to have behavioral biases and indicated that financial decisions can be driven by personality [5-7], level of sophistication [8], or even demographics [6, 9, 10].

For example, researchers [7] found that personality traits have some impact on an individual's risk-tolerance behavior, which ultimately influences investment decisions about stocks, securities and bonds. Findings by researchers [6] suggested that the fund managers' personality traits and demographics played a role in distinguishing high financial performance from average to poor financial performance. The personality traits used in their study are in accordance with Big-Five personality consisted of five basic dimensions, namely, Agreeableness (A), Conscientiousness (C), Extraversion (E), Neuroticism (N) and Openness to Experience (O) [11, 12]. Another's results [5] also exhibit that there is a direct correlation between investors' personality and financial decision bias. Furthermore, the characteristics of successful foreign exchange trader are found to be different depending on work location, type of foreign exchange instrument, and trading area [13].

This research differs from previous studies that focus on personality and demographics of the investors. In contrast, this study focuses on investing characteristics of individuals. Specifically, this paper explores the effects of investing characteristics on their performance in equity markets, particularly in the Stock Exchange of Thailand (SET), which is Thailand's national stock exchange. SET consists of 549 listed companies with a combined market capitalization of USD538.94 billion as of May 2019. The study first establishes comprehensive factors representing investors' characteristics, then tests whether these factors can explain differences in financial performance success of each investor. We found that investing characteristics have significant relationships with investors' financial performance.

The remainder of the paper is organized as follows: section 2 describes the methodology applied in this research, as well as data selection and sample description. The results are presented in section 3 and the discussion of the results are provided in section 4. Lastly, section 5 concludes.

## II. METHODOLOGY

In order to achieve our research objective, this exploratory study employed a survey-based method. Self-developed questionnaires were used to gather characteristics based on a preselected list of traits, which in turn is based on characteristics of famous and successful investors. Afterwards, exploratory factor analysis (EFA) was used to extract insight from the obtained questionnaires. Finally, relationships were tested using multiple linear regression.

Questionnaires were distributed to individual investors who currently invest in SET. The questionnaire consists of three parts in total.

1) *Investing characteristics*: In the first part, the items are related to investing characteristics. The items are based on characteristics of famous and successful investors, namely, Benjamin Graham, Philip Fisher, Warren Buffett and Charlie Munger, Sir John Templeton, George Soros, Peter Lynch, John Neff, and Anthony Bolton, resulting in a total of 84 items. The items are measured by a 5-point scale to the extent they agree with the statement defining each characteristic, from 1 = strongly disagree to 5 = strongly agree.

2) *Role model investor admiration*: The second part of the questionnaire measure how much the respondents admire each investor. The respondents were asked to indicate by a 6-point scale the extent to which they agree with the statement "I admire…": "Benjamin Graham," "Philip Fisher," "Warren Buffett and Charlie Munger," "Sir John Templeton," "George Soros," "Peter Lynch," "John Neff," and "Anthony Bolton," with 1 as "strongly not admire," to 5 as "strongly admire," and 6 as "not familiar."

3) *Respondents information*: The third part is relevant to respondents' investing information and their sociodemographic. They were asked about their investment style, experience in the market, industries currently investing in, returns and concurrent market returns. And they are ultimately asked about their gender, age, and education level.

III. RESULTS

A. Descriptive Statistics

A total of 403 questionnaires were returned. Table I reports descriptive statistics of the respondents. The majority (92.4%) of the respondents are male, while only 7.6% are female. Most respondents are either students or at the early stage of their careers since 52.7% are at the age of 18 to 30, 39.7% are 31 to 42, 6.8% are 43 to 54, and only 0.8% are above 55. As for the level of education, 6.4% are lower than bachelor's degree, 56.4% have a bachelor's degree, 34.6% have a master's degree, and 2.6% have a doctoral degree. Lastly, 60% of the respondents have been investing for no more than 6 years, 34.5% have 6 to 10 years of experience, 4% and 1.5% for 11 to 15 years and above 15 years, respectively.

TABLE I
DESCRIPTIVE STATISTICS

| Variables | Minimum | Maximum | Mean | SD |
|---|---|---|---|---|
| Age | 18.00 | 63.00 | 31.98 | 12.33 |
| Investing experience | 1.00 | 20.00 | 4.37 | 3.53 |
| Excess returns | -102.66 | 287.86 | 2.02 | 27.53 |

B. Exploratory Factor Analysis

83-item questions about investing characteristics were analyzed by EFA. After an iterative process of factor analysis, a 16-factor solution resulted. However, three of the factors appear to have high loadings on only one item each, thus they are disregarded. The final model consists of 13 factors, each with eigenvalues greater than 1, KMO measure of sample adequacy of 0.939, and sums of squared loading at approximately 60%, which is suggested to be an acceptable variance explained in factor analysis [14].

The names determined to represent best the investing characteristics subsumed in each of the 13 factors were middle/long time horizon (you are not investing short term by anticipating market direction; you are not interested in the market's short term changes), qualitative analyst (the stock you are holding now consists of competent managements and favorable competitive advantage; you do thorough research on the company before buying its stock), open-minded/disciplined (you are generous towards others; you are always patient, have control of self, and ethical), organized (you record your investing history; you make a summary record of the reasons you are buying a stock), emotional stability (you never buy or sell a stock because you are running out of patient; you never buy or sell a stock only because most people do), naïve (you do not expect a lot of returns; you buy a stock of the company you work at or you are used to), growth stock (the stocks you are holding are companies with high growth prospect and still unknown to other investors; the stocks you are holding are growing companies that you bought when it was cheap), concentrated portfolio (you do not diversify a lot; you do not diversify at all), contrarian (you do not follow traditional strategy of investing; you do not believe in economists' analysis on current economic outlook), value stock (you are successful because you always found stocks with appropriate price; you always sell stocks when its fundamental shifts from good to bad), globalized (you have invested in foreign stocks if it's price is attractive; you invest in various instruments and countries), intrinsic value (you always consider the firm's current asset when evaluating its stock; you always find a stock's intrinsic value using discounted cash flow), and price-independent (you do not use financial formula for buy/sell decision; even if the price has increased a lot, you still would not sell if you expect the business still has more value).

C. Effects of Investing Characteristics on Excess Returns

A multiple linear regression of excess returns on factors representing investing characteristics was conducted. Excess return is calculated by subtracting market return from each investor's actual return. We pre-analyzed the data for satisfaction of regression assumption to ensure that the estimates would be accurate. Scatter plot of predicted values of dependent variable and regression residuals show no patterns, indicating homoscedasticity. Sample size is sufficiently larger than

200, thus the regression is robust to violation of normality assumption. Autocorrelation is also of no concern since the data is not time series. Lastly, multicollinearity should not be present since the factors were derived using orthogonal rotation.

Model 1 in Table II presents estimates of the regression. Significant F-test (F (13,356) = 5.297, p < 0.000) and R-squared of 16.2% indicate that investing characteristics can adequately explain investors' excess return.

In model 1, individual t-tests show that open-mindedness/disciplined is positively related to excess returns (p < 0.10), organized has a significant positive effect on excess returns (p < 0.05), and emotional stability is positively related to excess returns (p < 0.10). These results suggest that investors who are more open-minded, disciplined, organized, and in control of their emotion tend to have a better performance. In contrast, naïve has a highly significant negative effect on excess returns (p < 0.01). This should come as no surprise since equity market is highly complex and dynamic, investors who do not have profound understanding of the market would not survive.

Growth stock and value stock are also positively related to excess returns and the relationship is highly significant (p < 0.000 and p < 0.001, respectively). These results indicate that these two competing strategies can outperform the market.

Concentrated portfolio has a highly significant positive effect on excess returns (p < 0.001), implying that investors with portfolios focusing on few stocks tend to have higher excess returns, this result is in contrast with modern portfolio theory in traditional finance where diversification can eliminate unnecessary idiosyncratic risks.

Lastly, contrarian is positively related to excess returns and the relationship is significant (p < 0.05), indicating that investors who are inclined not to follow the trend are more likely to obtain greater excess returns.

For robustness check, a multiple linear regression was run again with sociodemographic variables, we call this model 2. We used investing experience, age, and the level of education as additional control variables. Similar to model 1, pre-analysis was conducted for satisfaction of regression assumption. Scatter plot of predicted values of dependent variable and regression residuals show no patterns, suggesting there is no heteroscedasticity. Sample size is sufficiently larger than 200, thus the regression is robust to violation of normality assumption. Autocorrelation is also of no concern since the data is not time series.

TABLE II
REGRESSION RESULTS

| | Model 1: without demographics variables $F(13, 356)=5.297$, $R^2=0.162$, $\bar{R}^2=0.131$ | | | Model 2: with demographics variables $F(17, 319)=4.274$, $R^2=0.186$, $\bar{R}^2=0.142$ | | |
|---|---|---|---|---|---|---|
| | B | Std. Error | p-value | B | Std. Error | p-value |
| (Constant) | 1.577 | 1.178 | .182 | -10.986 | 7.793 | .160 |
| Middle/long time horizon | .810 | 1.180 | .493 | 1.324 | 1.294 | .307 |
| Qualitative analysis | .610 | 1.180 | .606 | -.193 | 1.313 | .883 |
| Open-minded/disciplined | 2.237 | 1.180 | .059 | 2.437 | 1.266 | .055 |
| Organized | 2.688 | 1.180 | .023 | 2.440 | 1.274 | .056 |
| Emotional stability | 2.027 | 1.180 | .087 | 2.447 | 1.299 | .061 |
| Naïve | -3.207 | 1.180 | .007 | -3.195 | 1.237 | .010 |
| Growth stock | 5.099 | 1.180 | .000 | 5.781 | 1.320 | .000 |
| Concentrated portfolio | 3.900 | 1.180 | .001 | 3.907 | 1.264 | .002 |
| Contrarian | 2.736 | 1.180 | .021 | 2.504 | 1.277 | .051 |
| Value stock | 4.073 | 1.180 | .001 | 4.000 | 1.262 | .002 |
| Globalized | -.419 | 1.180 | .722 | -1.174 | 1.274 | .357 |
| Intrinsic value | .300 | 1.180 | .799 | .322 | 1.292 | .804 |
| Price-independent | 1.617 | 1.180 | .171 | 1.764 | 1.271 | .166 |
| Gender | | | | 6.817 | 5.042 | .177 |
| Age | | | | -.085 | .105 | .416 |
| Education | | | | 2.040 | 2.118 | .336 |
| Investing experience | | | | .744 | .394 | .059 |

The results in Table II show that, similar to model 1, model 2 is significant (F (17,319) = 4.274, p < 0.000, R-squared = 0.186), and all significant individual t-tests are consistent with the previous results. Adjusted R-squared for model 2 (0.142) is also higher than model 1 (0.131), suggesting superior explanatory power of model 2 relative to model 1. Note that investing experience has a significant positive effect on excess returns (p < 0.10). This indicates that financial performance in stock market can be improved with experience. However, the level of education and age do not have a significant relationship with excess returns.

## IV. DISCUSSION

In this study, we first established comprehensive factors using the questionnaire of 84 items representing investing characteristics. This resulted in 13 main factors that together explain approximately 60% of the variance. Then we tested whether these factors can explain differences in financial performance success of each investor. Using multiple linear regression, we found that 8 of the 13 factors have significant relationship with excess returns. These factors are open-minded/disciplined, organized, emotional stability, naïve, growth stock, concentrated portfolio, contrarian, and value stock. The sign of coefficients for all significant estimates are positive except for naïve.

Note that there is a positive relationship between concentrated portfolio and excess return. This relationship indicates that investors with portfolios focusing on few stocks tend to have higher excess returns. This result is consistent with the previous researches, for example, [15], [16], [17]. However, it is in contrast to modern portfolio theory in traditional finance, which implies that the perfectly diversified market portfolio is the optimal investor's portfolio because of substantial benefits from diversification. Nonetheless, one possible explanation for this result is that based on information advantage theory, portfolios concentrated in a few markets and securities can be optimal for some investors because of information advantage they possess. Furthermore, though not significant, it is worth noting that being globalized is negatively related to excess returns. Being globalized could be perceived as one way of diversifying a portfolio, which is the opposite of concentrated portfolio. Consequently, this further supports the result that concentrated portfolio positively affects excess returns.

In addition, contrarian is positively related to excess returns, implying that investors who are inclined not to follow the trend are more likely to obtain greater excess returns. This finding is in line with previous study [18], which found that contrarian funds outperform herd funds by an average of 2.6% a year in the United States, and the other study [19], which found that contrarian mutual funds outperformed herd mutual funds by 0.5% per quarter in South Africa.

For robustness, further analysis was conducted using multiple linear regression with the same set of factors in addition of demographics, which are investing experience, age, and the level of education. The results are identical in terms of both sign and significance level, 8 of 13 factors representing investing characteristics exhibit relationships with excess return.

Furthermore, investing experience has a significant positive effect on excess returns. Since investing experience is a possible indicator of investor's level of sophistication [8], which reflects his/her expertise using financial instruments and how well he/she knows the financial market. Therefore, the level of investing experience is likely to have a positive relationship with excess return.

However, the level of education does not have a significant relationship with excess returns. This could be because level of education is also a possible indicator to identify sophisticated investors [20], however, since the level of investing experience is likely to be a more direct indicator of investor's level of sophistication, therefore, the effects are likely already captured resulting in a non-significant estimate for the level of education. Also note that the coefficient is still positive.

Lastly, though not significant, the findings indicate that high financial performance decreases with age. This might seem contradicting since age should be correlated with investing experience. However, this finding is actually in line with previous studies for example, [6], [9], and [10]. One study [9] suggested that although investors acquire skill with experience, pace and stress of the trading environment of trading is not considered a suitable environment beyond the age of 35 or 40. Another explanation is in the context of fund managers. Younger fund managers are more likely to have greater performance because younger individuals have stronger goal orientation [6], and they need to work harder because they not only have to advance in their careers, but are also more likely to be fired for their low performance [10].

## V. CONCLUSION

In this study, we derived the 13 factors representing investing characteristics using EFA on 84 items in the questionnaire. With multiple linear regression technique, we found that investing characteristics can explain differences in investors' financial performance. In particular, it is expected that investors with open-minded/disciplined, organized, emotional stability, growth stock, concentrated portfolio, contrarian, value stock, and investing experience are more inclined to have higher excess returns, while naïve investors are more likely to have lower excess returns. These results deepen our knowledge on what sort of investing characteristics are required to survive in the equity markets.